\documentstyle[amssymb,aps,preprint]{revtex}
\begin{document}
\draft
\preprint{CALT-68-2070}
\title{Bose-stimulated scattering off a cold atom trap}
\author{H. David Politzer}
\address{California Institute of Technology\\
Pasadena, California 91125\\
{\tt politzer@theory.caltech.edu}}
\date{August 28, 1996}
\maketitle

\begin{abstract}
The angle and temperature dependence of the photon scattering rate for
Bose-stimulated atom recoil transitions between occupied states is compared
to diffraction and incoherent Rayleigh scattering near the Bose-Einstein
transition for an optically thin trap in the limit of large particle number, 
$N$. Each of these three processes has a range of angles and temperatures
for which it dominates over the others by a divergent factor as $%
N\rightarrow \infty $.

\bigskip\ 
\end{abstract}

\pacs{PACS numbers 03.75.Fi, 42.50.Gy, 05.30.Jp, 32.90.+a}


The quantum statistics of trapped atoms can drastically effect small angle
light scattering from a cold, degenerate cloud of gas. The scattering of an
incident photon is accompanied by a transition between trap states of a
recoiling atom. At small enough scattering angle or momentum transfer, the
final atom state may very well already be occupied. For bosons there is an
enhancement of the transition rate proportional to $n_f+1$, where $n_f$ is
the number of bosons already in that final trap state. For fermions,
transitions to occupied states are blocked; the analogous factor is $n_f-1$.
(This Fock basis description is the natural one for the discussion of the
phenomenon, particularly in the context of gaseous systems.) Considerable
attention has been given to the consequent impact on resonant line-shape 
\cite{lewenstein} --- the scattered photon energy is shifted up and down by
the differences in trap-state energies, which are then averaged over the
contributing processes in the thermal ensemble. However, the enhancement of
the scattering rate itself and its angular dependence may also prove to be
dramatic and interesting effects \cite{javanainen,politzer1}.

Ref. \cite{politzer1} considered Born approximation scattering per unit
volume for a plane wave of light in a uniform medium of bosonic atoms as a
function of angle and temperature. As the temperature, $T$, drops below the
Bose-Einstein critical temperature, $T_c$, as a consequence of this Bose
enhancement the scattering rate increases sharply for angles corresponding
to momentum transfers ${\bf \mbox{\boldmath
$\delta$}}$ such that the number of particles per $(\hbar /\delta )^3$
volume is large. While plane waves and infinite, uniform media are
physically unattainable, they represent a reasonable approximation to a beam
that is many wavelengths wide incident on the central region of a cloud that
is many times yet larger still. This paper addresses another limit: when the
whole trap is illuminated uniformly by the incident beam. In particular, I
investigate the angle and $T$ dependence of the scattering rate for a plane
wave incident on an isotropic, harmonic trap containing $N$ non-interacting
atoms. Again, I consider only the Born approximation, i.\thinspace e., a
single scattering, which is appropriate to an optically thin trap. (This may
be realized by a combination of low density and sufficiently small basic
cross-section, e.\thinspace g., achieved by sufficient detuning from
resonance.) A useful reference level is given by the isotropic Rayleigh
scattering or fluorescence rate, and comparison with the even brighter
coherent diffraction at yet smaller angles is made. In addition to the
differential rates per solid angle, I also compare the total amounts of
scattered light due to the various processes, integrated over all angles.

There are many possible strategies for doing this conceptually quite
straightforward calculation. I have chosen to focus on the leading, large-$N$
behavior of each possible contributing sub-process for temperatures such
that $T/T_c\sim {\cal O}(1)$. In particular, while the trap ground state is
treated explicitly, the other states are described semi-classically,
e.\thinspace g., with a continuous density of states and stationary-phase
evaluations of WKB wave function overlaps. This allows analytic evaluation
(or at least estimates) of the various contributions to the total
scattering. It proves quite enlightening because the several sub-processes
and regions of scattering angle scale differently with $N$ at fixed $T/T_c$.
An alternative strategy of exact numerical evaluation would require
considerable detail over a great range in $N$ to extract comparable
information.

Laboratory detection of scattered light at very small angles has already
been achieved in the context of cold atom traps. Dramatic images of
condensation into a trap ground state have been obtained from coherently
diffracted light \cite{ketterle}. Essentially, one diverges all small angle
light, blocks out the unscattered forward beam, and then focuses the
remainder. Both diffraction and Bose-enhanced scattering have the properties
that they are brighter (per unit solid angle) than the isotropic scattering
and they do not contribute to heating the gas. Hence, they both make
possible non-destructive observation by allowing a greatly reduced
illuminating intensity for a given detected signal strength. However,
diffraction is confined to smaller and smaller angles with increasing trap
size. Thus, for larger traps of the future, clean diffractive signals may be
harder to extract. In contrast, while dimmer than diffraction at the
smallest angles, the Bose-enhanced scattering contributes over a much larger
range of angles --- a range that increases with $N$ --- and for much of this
region is considerably brighter than Rayleigh scattering.

In an infinite, uniform medium, diffraction cannot be addressed. It is of
infinite intensity but relegated to zero angle. In contrast, the present
analysis allows an explicit comparison of diffraction and Bose-stimulated
scattering. This calculation also underscores the fact that the latter
occurs only to the extent that the recoil from a photon can knock an atom
from one occupied state to another and is proportional to the product of the
two occupation numbers. As will become apparent, as $N\rightarrow \infty $
only a vanishing fraction of the total number of particles actively
participate in the Bose enhancement. As a consequence, the total luminosity
from Bose-stimulated scattering, integrated over angle, is significantly
less than that from diffraction. On the other hand, diffraction reflects the
overall particle density, and its angular distribution scales inversely as
the size of the system. In contrast, the Bose-enhanced scattering depends on
the momentum of particles in the occupied states. This momentum and,
consequently, the typical scattering angle grow with increasing temperature.

This paper is organized as follows. Section I contains the thermal
preliminaries. Section II contains the kinematics and the reference Rayleigh
scattering. Section III discusses diffraction. Bose-stimulated scattering is
evaluated in section IV, first for transitions to and from the trap ground
state and then for transitions between excited states; these are the only
``new'' formulae of this paper, but they make most sense in the context of
and in comparison with the other material. The qualitative lessons are
summarized in section V.

\section{Thermal preliminaries}

It is convenient to use the natural units of the three-dimensional isotropic
harmonic oscillator. In particular, take 
\[
m=\omega _0=\hbar =1\text{ }, 
\]
where $m$ is the particle mass and $\omega _0$ is the oscillator fundamental
frequency. Also, measure temperature $T$ in energy units (i.\thinspace e.,
the trap level spacing) so that Boltzmann's constant $k_B=1$. I will use
Cartesian coordinates throughout the analysis. So the single particle
stationary states are labeled by a triplet of integers ${\bf m}%
=(m_x,m_y,m_z) $, and the energy of such a state can be taken to be $%
\varepsilon _{{\bf m}}=m_x+m_y+m_z$. (As is evident from the form of $%
\varepsilon _{{\bf m}}$, {\bf m} is {\it not} a vector with respect to
spatial rotations.)

Divide the total number of particles $N$ as $N=N_0+N_e$, where $N_0$ is the
expected number of particles in the ground state (also written as $\langle
n_0\rangle $) and $N_e$ is the expected number in excited trap states,
i.\thinspace e., the remainder. For $N\rightarrow \infty $, such a system
has a critical temperature given by 
\begin{equation}
T_c=N^{1/3}\zeta (3)^{-1/3}  \label{Tc}
\end{equation}
where $\zeta (3)\simeq 1.202$ is the Riemann Zeta function \cite{deGroot}.
(In eq.\thinspace (\ref{Tc}), and throughout, $T$ is measured in units of
the trap level spacing; hence, to see a transition at fixed, finite,
physical temperature, one must increase the trap size as $N\rightarrow
\infty $.) For $T\leq T_c$, $N_0$ is a macroscopic fraction of $N$: 
\begin{equation}
N_0=N\left( 1-\left( \frac T{T_c}\right) ^3\right) \text{.}  \label{N0(T)}
\end{equation}
This can be deduced from the following evaluation of $N_e$. One begins by
noting that the expected occupation of any state ${\bf m}$ is given by 
\[
\langle n_{{\bf m}}\rangle =\frac 1{e^{(\varepsilon _{{\bf m}}-\mu )/T}-1}%
\text{ ,} 
\]
where $\mu $ is the chemical potential. It follows that 
\[
e^{-\mu /T}=1+\frac 1{N_0}\text{ .} 
\]
Hence, for all excited states we can take $\mu /T\simeq 0$ for $T\leq T_c$.
Thus, for $T\leq T_c$, 
\begin{eqnarray}
N_e &=&\int_0^\infty \frac{\frac 12m^2dm}{e^{m/T}-1}  \nonumber \\
&=&T^3\zeta (3)\text{ ,}  \label{Ne(T)}
\end{eqnarray}
which, by consistency, implies eq.\thinspace (\ref{N0(T)}). The integral in
eq.\thinspace (\ref{Ne(T)}) is obtained by approximating the sum over states 
${\bf m}$, whose degeneracy is $(\varepsilon _{{\bf m}}+1)(\varepsilon _{%
{\bf m}}+2)/2$ for large $N$ by an integral over $\varepsilon _{{\bf m}}$
(now called $m$). One can also deduce that imposing $N_e=N$ for $T\geq T_c$
implies 
\[
\frac{\mu _{T\geq T_c}}T\simeq -\frac{18\zeta (3)}{\pi ^2}\frac{(T-T_c)}{T_c}%
+{\cal O}\left( \left( \frac{T-T_c}{T_c}\right) ^2\right) \text{ .} 
\]

The rate for a boson transition from a particular initial state $i$ to a
particular final state $f$ is proportional to $n_i(n_f+1)$, where $n_i$ and $%
n_f$ are the actual occupation numbers. It is this product which then must
be thermally averaged if the system as a whole is characterized by an
equilibrium temperature. Hence, in principle, one needs to know the
two-level correlations $\langle n_in_f\rangle $ as well as the occupation
expectations, $N_i=\langle n_i\rangle $. For the purposes of the present
calculations, it suffices to use the replacement $\langle n_in_f\rangle
\simeq N_iN_f$ because the corrections to this form lead to changes in any
final result that are down by ${\cal O}(1/N)$. (Here, ``final result''
refers to the result {\it after} the thermal average over all possible
initial and final $N$-atom states is performed.) The reasons are discussed
at length elsewhere \cite{politzer2}, but, briefly, the issues are as
follows. For $i=f\neq 0$, the fractional fluctuations in the occupation of
any substantially occupied state are large (i.$\,$e., ${\cal O}(1)$).
However, for fixed $N$, the different $i$'s fluctuate almost independently.
So, for example, the variance associated with $N_e$ vanishes relative to $%
N_e $ as $N_e\rightarrow \infty $. Also, for $i\neq f$, the fixing of $N$
induces a correlation between occupations, but it is likewise negligible as $%
N\rightarrow \infty $. Finally, for $i=f=0$, fixing $N$ ensures that the
fluctuations in $n_0$ are microscopic, even for an ideal gas with $T\leq T_c$%
.

\section{General kinematics and Rayleigh scattering}

A photon of momentum ${\bf k}_i$ is incident on a trap. The goal is to
estimate the differential cross-section $d\sigma /d\Omega $ for scattering
the photon to momentum ${\bf k}_f$ directed in the solid angle $\Omega $
(and also the total cross-section, $\sigma $). In particular, I integrate
over the final photon frequency (or, equivalently, over $k_f$). In the
natural units defined above, the unit of velocity is of order the RMS
zero-point velocity of an atom in the trap ground state. In these units, the
speed of light, $c$, is very large. If $\Delta \varepsilon $ is the change
in the struck atom's energy in a particular collision and, therefore, minus
the change in the photon's energy, then $k_f=k_i-\Delta \varepsilon /c$.
Hence, the scattering cross-section $d\sigma /d^3{\bf k}_f$ is approximately
proportional to $\delta (k_f-k_i)$. If the final goal is angular dependence,
integrated over final photon energy, one can use the virtual equality of $%
k_i $ and $k_f$ to simplify the kinematics.

Let the momentum transfer be defined as ${\bf \mbox{\boldmath $\delta$}}=%
{\bf k}_f-{\bf k}_i$. At small angles, using $k_i=k_f$, $\delta $ is
proportional to the scattering angle $\theta $%
\[
\theta \simeq \frac \delta {k_i} 
\]
and 
\begin{equation}
d\Omega \simeq \delta \,d\delta \,d\varphi \,/\,k_i^2\text{ ,}
\label{dOmega}
\end{equation}
where $\varphi $ is the azimuthal angle. No information is lost by making
the trivial simplification of ignoring the effects of photon polarization on
angular dependence. (At issue is polarization in the scattering plane at
large angles; if this were of interest, it could easily be accounted for
correctly.) With this simplification, the photon angular dependence is $%
\varphi $-independent and totally determined by the magnitude of the
momentum transfer to the atom, $\delta $. Conversion to actual laboratory
scattering angles involves knowing the incident photon momentum or
frequency. However, the discussion of atomic trap physics is simplest and
clearest in terms of $\delta $.

Born approximation scattering separates naturally into three types of
processes: diffraction, Bose-stimulated scattering, and Rayleigh scattering.
By ``diffraction'' I mean the collisions in which the struck atom remains in
its original trap state. Hence, the $N$-atom final state is the same as the $%
N$-atom initial state, irrespective of which atom is struck. Therefore, the
amplitudes for photon scattering off each of the $N$ atoms must be added
coherently, and then the sum is squared to get the rate. When the struck
atom changes state, even if unobserved, the rates for all the different
possible processes are added (incoherently). As mentioned above, these rates
are proportional to $n_i(n_f+1)$. The $n_i\cdot 1$ piece is what is
traditionally identified as Rayleigh scattering. It is independent of the
occupation of the final state. The $n_i\cdot n_f$ piece is what I mean by
Bose-stimulated scattering. It is an analog for bosonic atoms of stimulated
emission for photons.

In all cases, the key transition matrix element that determines the
scattering connecting the trap state $i$ to $f$ is $\langle i|e^{i%
\mbox{\boldmath \footnotesize $\delta$}{\bf \cdot r}}|f\rangle $. This can
be visualized in momentum space as the overlap of the initial momentum wave
function shifted by the momentum transfer ${\bf \mbox{\boldmath $\delta$}}$
with the final state momentum wave function. However, many of the
calculations are easiest to evaluate in position space.

The problem at hand is separable, e.\thinspace g., in Cartesian coordinates.
Hence, the stationary state wave functions $\langle {\bf r|}i\rangle $ are
products of three one-dimensional harmonic oscillator wave functions.
Furthermore, the potential and the thermal distribution are isotropic. For
convenience, we can choose one of our axes, e.\thinspace g., {\bf \^{x}}, to
point along ${\bf \mbox{\boldmath
$\delta$}}$, and then 
\begin{equation}
\langle {\bf m|}e^{i\mbox{\boldmath \footnotesize $\delta$}{\bf \cdot r}}|%
{\bf m}^{\prime }\rangle =\langle m_x|e^{i\delta x}|m_x^{\prime }\rangle
\,\delta _{m_y,m_y^{\prime }}\,\delta _{m_z,m_z^{\prime }}\text{ .}
\label{mm'}
\end{equation}

The one-dimensional wave functions of interest are the ground state, 
\[
\langle x|0\rangle =\pi ^{-1/4}e_{}^{-x^2/2}\text{ ,} 
\]
and the WKB wave functions for $m\gg 1$. For example, for $x>0$%
\[
\langle x|m\rangle \simeq \left( \frac \pi 2\right) ^{-1/2}\left(
2m-x^2\right) ^{-1/4}%
{\;\sin \left( \int_x^a\sqrt{2m-x^2}dx+\frac \pi 4\right) \text{ for }x<a\; \atopwithdelims\{\} \;\frac 12\exp \left( -\int_a^x\sqrt{x^2-2m}dx\right) \text{ for }x>a\;}
\]
where $a=\sqrt{2m}$ is the classical turning point.

Rayleigh scattering serves as a convenient reference comparison for the
other processes. The Rayleigh differential cross-section is the total
Rayleigh cross-section for a single atom times the following expression: 
\[
\sum_{i,f}N_i\left| \langle i|e^{i\mbox{\boldmath \footnotesize $\delta$}%
{\bf \cdot r}}|f\rangle \right| ^2=N\text{ .} 
\]
Hence, Rayleigh scattering is independent of $T$ and ${\bf 
\mbox{\boldmath
$\delta$}}$ (i.\thinspace e., isotropic) and proportional to $N$.

For the sake of simplifying all formulae and facilitating the comparisons of
the various processes, I suppress the factor of the one-particle
cross-section and write all differential and total cross-sections in units
of that fundamental area. Hence, I write 
\begin{eqnarray*}
\frac{d\sigma _{\text{Rayleigh}}}{d\Omega } &=&N \\
\sigma _{\text{Rayleigh}} &=&4\pi N\text{.}
\end{eqnarray*}
To reiterate, the displayed ``cross-sections'' are not areas in the natural
trap units; rather, they all have a single, common factor, the atomic
physics one-particle cross-section, suppressed.

\section{Diffraction}

The diffractive cross-section is given by 
\[
\frac{d\sigma _{\text{diffraction}}}{d\Omega }=\left| \sum_iN_i\langle i|e^{i%
\mbox{\boldmath \footnotesize $\delta$}{\bf \cdot r}}|i\rangle \right| ^2%
\text{ .} 
\]
The $i^{\text{th}}$ term in the sum is just the Fourier transform of the
position space density of the $i^{\text{th}}$ one-particle state, weighted
by the occupation number. Hence, Born approximation diffraction has a very
classical interpretation, independent of the particle statistics: it is
always given by the square of the Fourier transform of the density. (Of
course, that density and its thermal behavior may themselves be very
dependent on quantum statistics.)

For large $N$ and $T\sim {\cal O}(T_c)$, it is appropriate to single out the
ground state and treat the excited states semi-classically. In particular,
the normalized position-space density for classical one-particle states of
energy $m$ in a three dimensional isotropic harmonic potential is 
\[
\rho _m({\bf r})=\frac 1{\pi ^2m^2}\sqrt{2m-r^2}
\]
for $r^2\leq 2m$ and $0$ otherwise. And the ground state density is just 
\[
\rho _0({\bf r})=\pi ^{-3/2}e^{-r^2}\text{ .}
\]
So, the large-$N$ diffractive cross-section is 
\begin{eqnarray}
\frac{d\sigma _{\text{diffraction}}}{d\Omega } &=&\left| \int d^3{\bf r\,}%
e^{i\mbox{\boldmath \footnotesize $\delta$}{\bf \cdot r}}\left[ N_0\rho _0(%
{\bf r})+\int_0^\infty \frac{\frac 12m^2dm}{e^{(m-\mu )/T}-1}\rho _m({\bf r}%
)\right] \right| ^2\text{ }  \nonumber \\
&=&\left| N_0e^{-\delta ^2/4}+\frac{4T}{\delta ^4}\int_0^\infty \frac{dz}{z^3%
}e^{-1/z}e^{\delta ^2\mu z/2}\right| ^2  \label{BesselK} \\
&=&\left| N_0e^{-\delta ^2/4}+\frac{4T}{\delta ^4}\right| ^2\text{ for }\mu
=0\text{ .}  \label{diff}
\end{eqnarray}
(The $z$ integral in eq.\thinspace (\ref{BesselK}) is an approximation to an
exact infinite sum that arises in the evaluation of the preceding line; the
integral representation is valid for $\delta ^2T\gg 1$.) Note that 
\[
\frac{4T}{\delta ^4}=\left( \frac 2{\delta ^2T}\right) ^2\frac{N_e}{\zeta (3)%
}\text{ .}
\]
Also, if the characteristic linear size of the diffuse cloud of atoms in
excited trap states at temperature $T$ is called $L$, then $L^2\sim {\cal O}%
(T)$; the corresponding, characteristic momentum transfer supported by
diffraction from such a cloud, $\delta _L$, would be ${\cal O}(1/L)$; so $%
\delta _L^2T\sim {\cal O}(1)$. Thus, the two terms in $d\sigma _{\text{%
diffraction}}/d\Omega $ have a clear interpretation. The diffraction
cross-section is nominally ${\cal O}(N^2)$; however, the $N_0^2$ part has
support only for $\delta \lesssim {\cal O}(1)$, i.\thinspace e., the inverse
of the spatial extent of the ground state. The excited state part has power
fall-off in $\delta $ and is only ${\cal O}(N_e^2)$ or larger for $\delta
\lesssim {\cal O}(T^{-1/2})\sim \,{\cal O}(N^{-1/6})$, where the latter
estimate applies for $T\sim {\cal O}(T_c)$. Of course, in the square of the
amplitudes, there is also a cross term between the two.

[For completeness, I note that for $\mu <0$ the $z$ integral in
eq.\thinspace (\ref{BesselK}) can be expressed as a sum of two ``modified
Bessel functions of the second kind'' of ranks 0 and 1 and argument $\delta 
\sqrt{-2\mu }$. However, the qualitative and asymptotic behavior is easiest
to extract from the integral representation itself.]

To integrate over angles, one must recall eq.\thinspace (\ref{dOmega}) for $%
d\Omega $. The excited-state part of the diffractive cross-section appears
to be singular as $\delta \rightarrow 0$. However, that is a reflection of
the inappropriateness of representing the discrete states as a continuum
when considering very small shifts in momentum. If one simply excludes the
vanishingly small forward cone as $N\rightarrow \infty $ in which the
diffuse cloud diffraction contributes, the remaining diffraction is
dominantly off the condensate: 
\begin{equation}
\sigma _{\text{diffraction}}\simeq \frac{2\pi }{k_i^2}N_0^2\text{ .}
\label{diff2}
\end{equation}
Alternatively, one can estimate where the discreteness of the spectrum would
cut off the integral. For excited states with energies of ${\cal O}(T)$, the
difference in RMS momentum from one level to the next is ${\cal O}(T^{-1/2})$%
. Using that as an estimate for the smallest meaningful $\delta $ in
integrating eq.\thinspace (\ref{diff}), one learns that the integrated
cross-section due to excited states alone is ${\cal O}(N_e^{5/3}/k_i^2)$,
which does not compete with eq.\thinspace (\ref{diff2}) for $T<T_c$ as $%
N\rightarrow \infty $.

The explicit appearance of the incident photon momentum $k_i$ is inevitable
here (and also below) when considering total cross-sections. While the
Rayleigh differential cross-section is genuinely isotropic, diffraction and
Bose-stimulated scattering are only significant at small angles. Within
those ranges they can be much larger than the Rayleigh rate, as indicated by
the formulae. However, the extent of those ranges is governed by the ratio
of the acceptable struck-atom momentum transfer (which depends on trap
geometry and temperature) to the incident photon momentum, which is
typically much larger.

\section{Bose-stimulated scattering}

The Bose-stimulated rate has two sources: scattering atoms into and out of
the condensate and scattering atoms between excited trap states. For each of
these classes of processes, one first determines the square of the matrix
element of $e^{i\mbox{\boldmath \footnotesize $\delta$}\cdot {\bf r}}$
between initial and final states. Again, it is convenient to use ${\bf %
\mbox{\boldmath $\delta$}}$ to define one of the Cartesian coordinate
directions, e.\thinspace g., ${\bf \mbox{\boldmath $\delta$}}=\delta {\bf 
\hat{x}}$. Then the relevant overlap is an integral over one-dimensional
oscillator wave functions. Also, only states which share the same {\bf \^{y}}
and {\bf \^{z}} quantum numbers contribute, as indicated in eq.\thinspace (%
\ref{mm'}). So the thermal sum over occupations of possible states involves
first a projection onto the contributing states and then a sum over the
thus-projected occupations weighted by the non-trivial one-dimensional
matrix-element-squared.

\subsection{In and out of the condensate}

With the choice ${\bf \hat{x}}={\bf \mbox{\boldmath $\delta$}/}\delta $,
only states labeled by ${\bf m}=(m,0,0)$ have non-zero matrix elements of $%
e^{i\mbox{\boldmath \footnotesize $\delta$}{\bf \cdot r}}$ with the ground
state. The one-dimensional overlap of relevance for transitions between the
trap ground and excited states is 
\[
F(m,\delta )\equiv \left| \langle 0|e^{i\delta x}|m\rangle \right| ^2\text{ .%
} 
\]
For large $N$ and $T\sim {\cal O}(T_c)$, only $m\gg 1$ is of interest. $F$
is sharply peaked around $\delta \approx \sqrt{2m}$, the maximum classical
momentum of a one-dimensional harmonic oscillator of energy $m$. Its width
in $\delta $ is ${\cal O}(1)$, which is vanishingly small on the scale of
the typical value of $\sqrt{m}$. Hence, it suffices to represent $F$ as a $%
\delta $-function, whose normalization can be deduced by integrating $F$
over $\delta $. The result is 
\[
F(m,\delta )\simeq \delta (m-\frac{\delta ^2}2)\text{ .} 
\]

This behavior has a simple classical interpretation. To bring an energetic
particle in a three-dimensional harmonic potential to rest with a single
impulse and have it remain at rest, the following must be true: the
energetic orbit must pass through the origin; the impulse must be applied
when the particle passes through the origin; and the impulse must be
precisely that required to bring the particle to rest.

The Bose-stimulated differential cross-section is obtained by summing over $m
$ the equal contributions of transitions $0\rightarrow m$ and $m\rightarrow 0
$: 
\begin{eqnarray*}
\frac{d\sigma _{\text{Bose}}^{0,m}}{d\Omega } &=&2N_0\int_0^\infty \frac{dm}{%
e^{m/T}-1}F(m,\delta ) \\
&=&\frac{2N_0}{e^{\delta ^2/(2T)}-1}\text{ } \\
&\approx &\;%
{\;2N_0e^{-\frac{\delta ^2}{2T}}\text{ \quad for }\delta ^2\gtrsim 2T\; \atopwithdelims\{\} \;4N_0\frac T{\delta ^2}\text{ \quad for }\delta ^2<2T\;}
\text{ .}
\end{eqnarray*}
For $\delta \sim {\cal O}(1)$ this is ${\cal O}(N_0N_e^{1/3})$ while the
diffractive cross-section is ${\cal O}(N_0^2)$; however, the latter then
falls off exponentially with $\delta ^2$ while the $0\leftrightarrow m$
stimulated rate only falls as $1/\delta ^2$. Note also that this stimulated
rate is much larger than the Rayleigh rate as long as $\delta <{\cal O}%
(T^{1/2})$. When $\delta ^2>T$, the $0\leftrightarrow m$ stimulated rate
falls off exponentially with $\delta ^2/2T$ because of the fall-off in thermal
occupation of appropriately high momentum states.

This analysis is again inadequate for very small $\delta $, where the
discreteness of the spectrum is relevant. However, we can integrate down to $%
\delta \sim {\cal O}(1)$ for which these approximations are sufficient and
estimate 
\[
\sigma _{\text{Bose}}^{0,m}(\delta \gtrsim 1)\simeq \frac{4\pi N_0T}{k_i^2}%
\log (2T)\text{ .} 
\]

\subsection{Between excited states}

The excited states ${\bf m}$ and ${\bf m}^{\prime }$ must be related by $%
{\bf m}=(m,m_y,m_z)$ and ${\bf m}^{\prime }=(m^{\prime },m_y,m_z)$ to get a
non-zero overlap with $e^{i\mbox{\boldmath \footnotesize $\delta$}{\bf \cdot
r}}$. The relevant one-dimensional matrix element is 
\[
G(m,m^{\prime },\delta )\equiv \left| \langle m|e^{i\delta x}|m^{\prime
}\rangle \right| ^2\text{ .} 
\]
This is to be summed against the thermal occupations projected onto the one
dimension: 
\begin{eqnarray*}
{\cal P}(m,m^{\prime },T,\mu ) &\equiv &\sum_{m_y,m_z}\left[
(e^{(m+m_y+m_z-\mu )/T}-1)(e^{(m^{\prime }+m_y+m_z-\mu )/T}-1)\right] ^{-1}
\\
&=&T^2P\left( \frac{m-\mu }T,\frac{m^{\prime }-\mu }T\right)
\end{eqnarray*}
where 
\begin{equation}
P(a,b)\equiv \int_0^\infty \frac{z\,dz}{(e^{z+a}-1)(e^{z+b}-1)}\text{ .}
\label{P(a,b)}
\end{equation}
Thus, in terms of the functions so defined, 
\[
\frac{d\sigma _{\text{Bose}}^{m,m^{\prime }}}{d\Omega }=\int dm\,dm^{\prime
}\,G(m,m^{\prime },\delta )\,{\cal P}(m,m^{\prime },T,\mu )\text{ .} 
\]

$P(a,b)$ can be expressed in closed form in terms of exponentials,
logarithms, and dilogarithms. Again, though, its qualitative and limiting
behavior is easiest to see from the defining integral, eq.\thinspace (\ref
{P(a,b)}). $P(a,b)$ is symmetric in $a$ and $b$, and it is ${\cal O}(1)$
when both $a$ and $b$ are themselves ${\cal O}(1)$. If $b\leq a\ll 1$, it is 
${\cal O}(\log (1/a))$. And it is ${\cal O}(\exp (-a-b))$ when $a,b>1$.
(Recall, also, that $\mu /T$ is $0$ for $T\leq T_c$ and ${\cal O}%
((T-T_c)/T_c)$ for $T\gtrsim T_c$.)

$G(m,m^{\prime },\delta )$ can be evaluated by stationary phase using the
WKB wave functions. For each $m$, $m^{\prime }$, and $\delta $ (all three
positive), there is a single position-space point of stationary phase in the
overlap integral, corresponding to the unique classical position where an
impulse of magnitude $\delta $ can change a particle of energy $m$ to one of
energy $m^{\prime }$. $G(m,m^{\prime },\delta )$ is symmetric in $%
m,m^{\prime }$; the following simple form is for $m\geq m^{\prime }$: 
\[
G(m,m^{\prime },\delta )=%
{\;\frac 1{2\pi }[2m^{\prime }\delta ^2-(m-m^{\prime }-\frac{\delta ^2}2)^2]^{-1/2}\; \atopwithdelims\{\} \;\text{or }0\text{ if }2m^{\prime }\delta ^2<(m-m^{\prime }-\frac{\delta ^2}2)^2\;}
\text{ .} 
\]
Unlike $F(m,\delta )$, the support of $G(m,m^{\prime },\delta )$ for $%
m,m^{\prime }\gg 1$ extends over a non-negligible region in $\delta $
relative to the typical thermal values of $m$ and $m^{\prime }$. $%
G(m,m^{\prime },\delta )$ has an integrable square-root singularity on its
boundary in the $m$-$m^{\prime }$ plane. That singularity occurs when the
point of stationary phase is the origin in position space. The stationary
phase evaluation takes account not only of how likely it is to find the
particle at the point of stationary phase but also how rapidly the phase
begins to change as you go away from that point. It is this latter aspect
that is optimized by applying the impulse when the particle is at the
origin, and that is where the maxima of $G(m,m^{\prime },\delta )$ occur.

Combining these factors, one finds, e.\thinspace g., for $T\leq T_c$, 
\begin{eqnarray}
\frac{d\sigma _{\text{Bose}}^{m,m^{\prime }}}{d\Omega } &=&2\int_{\delta
^2/8}^\infty dm\int_{(\sqrt{m}-\delta /\sqrt{2})^2}^mdm^{\prime
}\,G(m,m^{\prime },\delta )\,{\cal P}(m,m^{\prime },T,0)  \nonumber \\
&=&\frac{T^3}\pi \int_{a/4}^\infty dx\int_{x+a-2\sqrt{ax}}^xdy\int_0^\infty
z\,dz\{(e^{z+x}-1)^{-1}(e^{z+y}-1)^{-1}\cdot   \nonumber \\
&&\hspace{0.8in}\cdot (y-x-a+2\sqrt{ax})^{-1/2}(x+a+2\sqrt{ax}-y)^{-1/2}\}
\label{m,m'}
\end{eqnarray}
where 
\[
a=\frac{\delta ^2}{2T}\text{ .}
\]
Hence, $d\sigma _{\text{Bose}}^{m,m^{\prime }}/d\Omega $ is $T^3$ (remember
that $T\simeq N_e^{1/3}$) times a function $f$ of the ratio $a$ defined
above. When $a\sim {\cal O}(1)$, $f(a)$ is ${\cal O}(1)$. When $a\gg 1$, $%
f(a)$ is ${\cal O}(\exp (-a/4))$. And, when $a\ll 1$, $f(a)$ is still ${\cal %
O}(1)$. To ascertain this last feature, note that the range of integrated $y$
vanishes as $a\rightarrow 0$; $y$ can be estimated as $x$ in the $z$
integral, yielding a $\log (1/x)$ integrable singularity upon integrating $z$%
, while the singularities in the $y$ integral itself are also integrable,
yielding approximately $\pi /2$. Hence, the apparent divergence of the $z$
integral (as $x,y\rightarrow 0$) is not realized, even for $a\rightarrow 0$.
A summary of these estimates can be represented as 
\[
\frac{d\sigma _{\text{Bose}}^{m,m^{\prime }}}{d\Omega }\sim {\cal O}\left(
N_ee^{-\frac{\delta ^2}{8T}}\right) 
\]
for all $\delta $.

This is the crudest of the explicit determinations of the coherent processes
(simply because eq.\thinspace (\ref{m,m'}) is the furthest from a closed
form in terms of elementary functions). However, it is also the only
coherent process whose differential rate never exceeds the Rayleigh rate by
a divergent factor as $N\rightarrow \infty $. So there is no range of angle
or temperature for which it is the overwhelmingly dominant process. A
numerical determination of $f(a)$ is certainly feasible, but the detailed
result would shed no light on the present, general discussion. It would be
needed, though, in the context of a particular trap potential for a
particular experiment if one wanted to account for all the light at each
angle to better than a factor of two or three.

From the estimated behaviors given above, one can estimate the $m,m^{\prime
} $ contribution to the cross-section: 
\[
\sigma _{\text{Bose}}^{m,m^{\prime }}\sim {\cal O}\left( \frac{N_e^{4/3}}{%
k_i^2}\right) \text{ ,} 
\]
which is down by a factor of $\log (T)$ relative to $\sigma _{\text{Bose}%
}^{0,m}$, at least when $N_0$ is comparable to $N_e$, because the
differential $m\leftrightarrow m^{\prime }$ cross-section does not grow like 
$T/\delta ^2$ as $\delta ^2$ decreases below $T$.

\section{Summary and conclusions}

I have chosen units and notation to emphasize the angle, temperature, and
number dependence of light scattering off an optically thin cold atom trap.
The angle is best represented in a general discussion by the momentum
transfer, ${\bf \mbox{\boldmath $\delta$}}$. I have focussed on
temperatures, $T$, near the Bose-Einstein condensation temperature, $T_c$.
And I have used semi-classical analytic techniques to evaluate the leading
large-$N$ (particle number) behavior of each type of contributing process.

Rayleigh scattering or ordinary fluorescence corresponds to the $n_i\cdot 1$
term in the factor $n_i(n_f+1)$ that characterizes boson transition rates.
Hence, the sum over all possible initial and final atom trap states gives a
factor of $N$ for the Rayleigh rate. Because all final states are weighted
equally, there is no dependence on ${\bf \mbox{\boldmath $\delta$}}$.

Diffraction reflects the particle density. The characteristic $\delta $ of
each sub-process is the inverse of the density's characteristic length
scale. In a harmonic potential, using natural units, the ground state has
size ${\cal O}(1)$. Thermally occupied states have size ${\cal O}(T^{-1/2})$%
. Hence, the most prominent (largest angle) feature of diffraction occurs
for $T<T_c$ due to scattering off the condensate. The condensate diffraction
rate is proportional to $N_0^2$ for the region $\delta \lesssim 1$ (in the
natural oscillator units used throughout). For larger $\delta $, this term
drops off with an $\exp (-\delta ^2/2)$ because there just is no more
probability to find condensate particles at larger momenta. The relevant $%
n_f $ is, indeed, $n_i-1$. One must also, in principle, add all possible
struck atoms coherently. However, the diffraction from the diffuse, thermal
cloud corresponding to excited trap states has support only for $\delta
\rightarrow 0$ as $N\rightarrow \infty $. Because diffraction is limited in $%
\delta $, in contrast to the $\delta $-independent Rayleigh scattering, the
diffraction contribution to the total cross section has a factor of $(\bar{%
\delta}/k_i)^2$, where $\bar{\delta}$ is the mean momentum transfer for the
process and $k_i$ is the initial photon momentum.

``Bose-stimulated scattering'' is used here to refer to the $n_i\cdot n_f$
term, where coherence among the $n_f$ bosons in the final state leads to an
enhancement of the rate relative to the totally incoherent Rayleigh rate.
This is a dramatic feature of gases of Bose condensed atoms for $T\lesssim
T_c$ when several of the lowest lying states have large occupation numbers.
The characteristic $\delta $ is of order the typical thermal particle
momentum. This momentum is ${\cal O}(T^{1/2})$. However, the sum over $i$
and $f$ does {\it not} give cross-sections proportional to $N_0N_e$ or $%
N_e^2 $, where $N_e=N-N_0$, even when integrated over all possible ${\bf %
\mbox{\boldmath $\delta$}}$. Instead, it was shown explicitly that only $%
{\cal O}(N_e^{1/3})$ of the $N_e$ particles in excited states have an
overlap with the ground state after absorption of the photon recoil
momentum. The reason that the $n_i$ (or $n_f$) factor does not lead to an $%
N_e$ in the final result is that the typical excited state particle has an
orbit or wave function that is inherently three dimensional. The 1/3 power
comes just from the reduction from the three-dimensional ensemble to the one
dimension of contributing, active participants. Transferring a particle from
one typical occupied excited state to another with a single impulse from the
photon proves to be, if anything, a bit harder. Again, only ${\cal O}%
(N_e^{1/3})$ of the $N_e$ possible final particles contribute.

The scattering rate from the $0\leftrightarrow m$ processes grows like $%
T/\delta ^2$ for angles corresponding to $\delta ^2\lesssim 2T$, and this $%
T/\delta ^2$ is the enhancement factor over the Rayleigh differential rate
in the corresponding angular region. The angular size of this region grows
with $N$, since $T\approx N_e^{1/3}$. This is in contrast to diffraction
from the condensate, which is limited to $\delta \sim {\cal O}(1)$. The $%
m\leftrightarrow m^{\prime }$ rate, though monotonically increasing with
decreasing $\delta $, is never substantially bigger than the Rayleigh rate.
Also, both Bose-stimulated processes fall off exponentially with $\delta ^2$
for $\delta ^2>2T$ again because of the absence of particles with higher
momenta.

The heart of the calculation of the Bose-stimulated processes is the
estimate of the efficacy of a photon momentum transfer, ${\bf %
\mbox{\boldmath $\delta$}}$, to effect a transition between occupied trap
states. The details of the transition matrix elements will depend on the
explicit trapping potential. Whatever the potential, the integration over
all ${\bf \mbox{\boldmath $\delta$}}$ is two- rather than three-dimensional
because of the constraint of overall energy conservation. The implies that
even integrating over all kinematically accessible ${\bf 
\mbox{\boldmath
$\delta$}}$ cannot produce a Bose stimulation factor that is a significant
fraction of $N$. Instead, it is generically a fractional power.

If one observes the photon energies with high resolution (instead of
integrating over them as done in the calculations presented here), there
will be a strong angular dependence to the line shape. In particular, the
different processes discussed here produce different photon spectra. At the
smallest angles, the dominant diffraction leaves the photon energy
unchanged. The Bose-enhanced scattering, dominant at somewhat larger angles,
occurs with photon energy shifts of ${\cal O}(T)$, while Rayleigh scattering
has a characteristic, larger shift due to atom recoil.

\end{document}